# DEEP LOW-LATENCY JOINT SPEECH TRANSMISSION AND ENHANCEMENT OVER A GAUSSIAN CHANNEL

*Mohammad Bokaei[1], Jesper Jensen[1,2], Simon Doclo[3], Jan Østergaard[1]*

[1] Aalborg University, Aalborg, Denmark    [2] Oticon A/S, Copenhagen, Denmark
[3] University of Oldenburg, Dept. of Medical Physics and Acoustics, Germany

## ABSTRACT

Ensuring intelligible speech communication for hearing assistive devices in low-latency scenarios presents significant challenges in terms of speech enhancement, coding and transmission. In this paper, we propose novel solutions for low-latency joint speech transmission and enhancement, leveraging deep neural networks (DNNs). Our approach integrates two state-of-the-art DNN architectures for low-latency speech enhancement and low-latency analog joint source-channel-based transmission, creating a combined low-latency system and jointly training both systems in an end-to-end approach. Due to the computational demands of the enhancement system, this order is suitable when high computational power is unavailable in the decoder, like hearing assistive devices. The proposed system enables the configuration of total latency, achieving high performance even at latencies as low as 3 ms, which is typically challenging to attain. The simulation results provide compelling evidence that a joint enhancement and transmission system is superior to a simple concatenation system in diverse settings, encompassing various wireless channel conditions, latencies, and background noise scenarios.

*Index Terms*— low-latency, joint source-channel coding, analogue speech transmission, speech enhancement.

## 1. INTRODUCTION

Speech transmission and enhancement in low-latency scenarios pose significant challenges in ensuring high intelligibility and quality communication [1, 2]. Traditional speech enhancement methodologies often rely on simple statistical models [3], while conventional speech transmission strategies encompass separate approaches for source coding and channel coding [2]. Recent advancements in deep learning have showcased promising outcomes in addressing both enhancement [4, 5] and speech transmission [6, 7] objectives.

In the domain of speech enhancement, data-driven approaches using deep learning have emerged as powerful alternatives to conventional methods, demonstrating superior performance in enhancing speech quality and reducing noise [4, 5, 8]. However, many data-driven speech enhancement methods are non-causal, resulting in high latencies, often exceeding a second [5, 9]. Researchers have proposed causal variants of these methods to prioritize lower latencies, albeit at the expense of some performance.

Traditional methods for low-latency speech transmission often rely on separate source-channel coding techniques [2]. While recent advances in deep learning-based source coding (audio codecs) [6, 10] and channel coding [11, 12] have shown superior performance over conventional methods, these approaches are typically developed independently, which may impact overall system performance and introduce extra latency [2]. In the context of low-latency speech and audio transmission, several researchers have explored deep learning techniques for joint source-channel coding [7, 13, 14]. However, very little focus has been given to handling noisy inputs effectively, i.e., low-latency joint speech enhancement and source-channel coding.

In the context of joint speech transmission and enhancement, the presence of acoustic background noise, speech coding, and communication over noisy wireless channels introduces sources of error in the clean speech signal. Traditional joint source-channel-based speech transmission techniques primarily address speech coding and transmission channel effects, while speech enhancement systems focus on mitigating background noise. However, only a few studies have attempted to tackle all these error sources simultaneously. One notable approach is the development of a noise-robust audio codec [15], which handles audio coding and background noise; however, it does not include transmission channel effects. Also, an investigation of all

This project has received funding from the European Union's Horizon 2020 research and innovation programme under the Marie Skłodowska-Curie grant agreement No.956369.

three error sources is presented in [16], where the authors assume digital communication and model communication channel effects as packet losses with a latency of 20 ms. Such high latency is impractical for low-latency applications, e.g., wireless hearing assistive devices [17].

In this paper, we propose novel solutions for low-latency joint speech analogue transmission and enhancement, harnessing the potential of deep neural networks. Analogue communication, with its lower bandwidth requirements and latency compared to digital communication, presents an attractive alternative in low-latency applications. Our approach combines Conv-TasNet [5] state-of-the-art DNN architectures for speech enhancement and TransNet, a state-of-the-art DNN architecture for analogue joint source-channel-based speech transmission [18]. The latency of the proposed system is configurable but can operate with total system latencies as low as 3 ms.

We seek to address the challenges posed by acoustic background noise, speech coding, and transmission channel errors simultaneously in the context of low-latency joint speech analogue transmission and enhancement. By incorporating deep learning-based methods, we aim to offer an innovative solution that effectively handles these error sources and ensures high-quality speech communication in real-time applications.

We propose two novel methods for low-latency joint speech transmission and enhancement. Firstly, we concatenate pre-trained versions of Conv-TasNet and TransNet to form a combined system. This is a simple solution for low-latency speech transmission and enhancement problems because it uses existing pre-trained DNNs and does not require further training. This problem has not been studied before to the best of our knowledge. Secondly, we jointly train both systems using an end-to-end approach to optimize their performance collaboratively. We conduct comprehensive simulations, comparing the performance of these methods for various transmission bandwidths, wireless SNRs, and system latencies. The simulations demonstrate the superiority of the joint method over the simple concatenation method in terms of estimated speech intelligibility and reconstruction quality.

## 2. LOW LATENCY SPEECH TRANSMISSION AND ENHANCEMENT

In this section, we present our low-latency speech transmission and enhancement system, comprising two distinct subsystems: one dedicated to speech enhancement and the other to speech transmission. We will begin by providing separate descriptions of the speech enhancement and speech transmission systems in the following subsections. Subsequently, we introduce our proposed method.

### 2.1. Low latency speech enhancement

For achieving low-latency speech enhancement, we employed the Conv-TasNet architecture, as proposed in [5]. This system deals with background noises as a source of error in the input speech signal. Fig. 1(a) provides an overview of its structure. In the Conv-TasNet system, input frames in the time domain are transformed into inner domain representations through the Encoder, which is a learned linear map. Subsequently, the Separator estimates scalar weight values which are applied point-wise to the inner domain representation to enhance the signal. An enhanced time-domain output waveform is then obtained through the Decoder, another learned linear map. In our work, we implement Conv-TasNet with causal convolutions and cumulative layer normalization (cLN), resulting in a latency of 3 ms of Conv-TasNet. The full detail of causal Conv-TasNet is provided in [5].

### 2.2. Low latency speech transmission

In the proposed joint low-latency speech enhancement and transmission system, we used the system from [18] for transmission, which we refer to as "TransNet" [18]. The TransNet structure is shown in Fig. 1 (b). This structure has demonstrated outstanding performance in various audio applications, including neural vocoders [19, 20], audio codecs [6, 10], and artificial bandwidth extension [21, 22], and recently for speech transmission [18].

TransNet is a deep analogue joint source-channel-based speech transmission approach with configurable latency ranging from 2-8 ms [18]. This system effectively addresses two sources of degradation in the transmitted speech. The first source pertains to distortion caused by audio coding, while the second source involves the effects of wireless transmission on the encoded data. The system is composed of an encoder, a decoder, and a model of a wireless transmission channel Fig. 1(b). The encoder commences with a 1D convolutional layer followed by three layers of encoder blocks with stride = 2, a 1D convolutional layer, and concludes with a layer normalization layer. These five layers have 4, 8, 16, 16, and 8 kernels, respectively. The encoder blocks consist of three residual units with dilations convolutions of 1, 3, and 9, followed by a convolution layer to apply the stride. The encoder downsampling rate is 8. The encoded data is transmitted over a wireless channel and subsequently received at the decoder. The decoder mirrors the encoder structure but in reverse order, excluding the normalization layer and substituting transpose convolutions with normal convolution layers for upsampling inside decoder blocks. Further details on the encoder and decoder blocks can be found in [6].

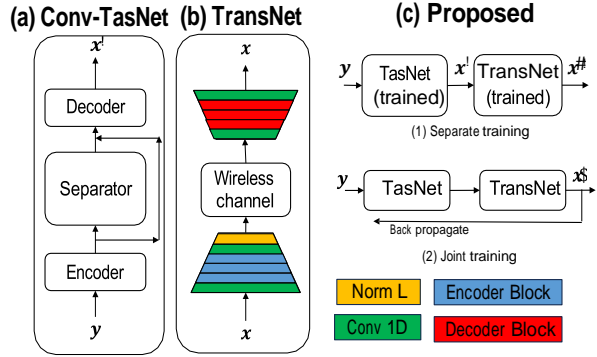

**Fig. 1**: Overview and the structure of the proposed methods, TransNet [18], and Conv-TasNet [5].

**2.3. Proposed methods**

We present two novel methods for achieving low-latency speech transmission and enhancement. In both methods, we leverage a low-latency speech enhancement module (Conv-TasNet) and a low-latency speech transmission module (TransNet) and cascade them. In the first approach, we separately train the Conv-TasNet and TransNet modules, each serving the specific independent purpose of speech enhancement and speech transmission, respectively. In contrast, the second method involves joint training of both systems. This approach aims to optimize the performance of both modules collaboratively, fostering a more integrated and efficient system for low-latency speech transmission and enhancement.

Conv-TasNet, in its default setting, contains 5.1M parameters, while TransNet contains 20k parameters. Conv-TasNet precedes TransNet in the processing sequence Fig. 1(c.1). This order is more suitable for applications where the noisy speech signal is captured in a device with high computational complexity available, e.g., a wireless microphone, and transmitted to a low-complexity device, e.g., a hearing assistive device.

*2.3.1. System 1: Separate training*

The proposed separate training method is illustrated in Fig. 1(c.1). We discuss the separate training of Conv-TasNet and TransNet separately in following.

**Conv-TasNet:** Let us assume the noisy input speech signal is represented as $\mathbf{y} = \mathbf{x} + \mathbf{n}_a$, where $\mathbf{x} \in \mathbb{R}^n$ denotes the clean speech signal and $\mathbf{n}_a \in \mathbb{R}^n$ represents the background acoustic noise. In this separate training context, the objective of the Conv-TasNet network is to take the noisy input $\mathbf{y}$ and produce an estimated speech signal, $\mathbf{x}' \in \mathbb{R}^n$ that closely resembles the clean speech signal $\mathbf{x}$, Fig. 1(a). To achieve this goal, the Conv-TasNet network is trained using the Scale-Invariant Signal-to-Distortion Ratio (SI-SDR) metric denoted by $\mathcal{L} = $ SI-SDR$(\mathbf{x}, \mathbf{x}')$ as the cost function [5].

**TransNet:** During the separate training of TransNet, we consider a scenario where a clean speech signal $\mathbf{x}$ is provided, Fig. 1 (b). The objective is to obtain an output speech signal $\tilde{\mathbf{x}} \in \mathbb{R}^n$ from the decoder closely resembling the original input speech signal. We model the wireless communication channel as Additive White Gaussian Noise (AWGN) (see Fig. 1(b)), where noise $\mathbf{n}_w \in \mathbb{R}^k$ is added to the encoded signal, with $k$ representing the dimension of the encoded signal generated by TransNet. To train TransNet, we utilize the Mean Square Error (MSE) metric as the cost function, given by $\mathcal{L} = $ MSE$(\mathbf{x}, \tilde{\mathbf{x}})$. In addition, we introduce a parameter called the bandwidth compression ratio, denoted as $R = k/n$, which represents the ratio between the dimension of the input speech signal frame ($n$) and the dimension of the encoded data ($k$) in TransNet [18]. The bandwidth compression ratio allows us to quantify the level of data compression achieved during the speech transmission process.

*2.3.2. System 2: Joint training*

Fig. 1(c.2) presents the overview of our proposed joint training method. The system takes the noisy speech signal $\mathbf{y}$ as input and aims to jointly perform speech enhancement and transmission, yielding the output signal $\hat{\mathbf{x}}$. The first stage of the proposed

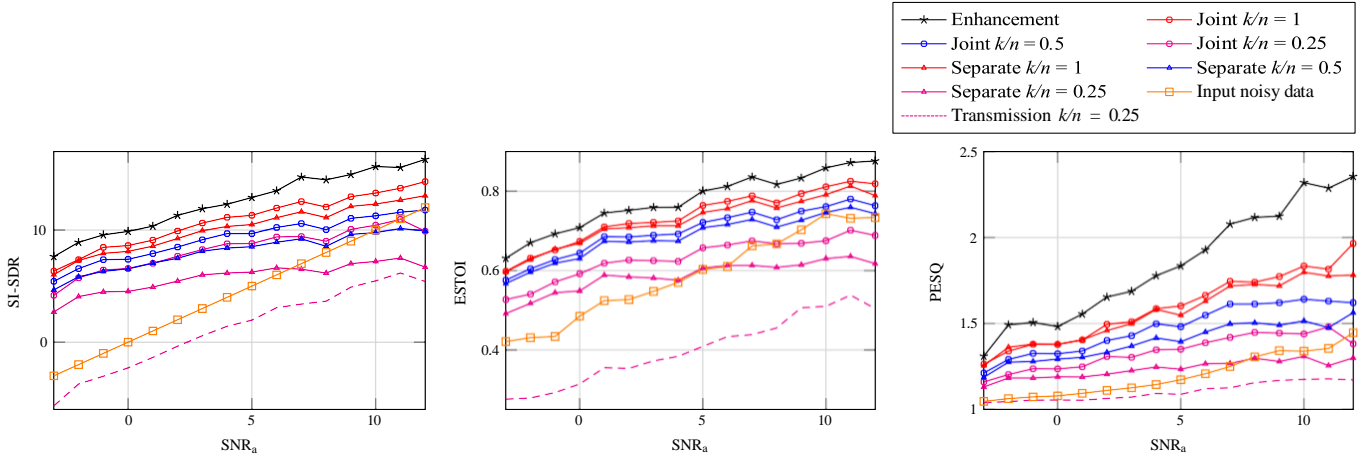

**Fig. 2**: Performance of the proposed joint enhancement and transmission methods for different bandwidths in terms of speech intelligibility (ESTOI), speech quality (PESQ), and reconstruction quality (SI-SDR) versus $SNR_a$ of the speech signal against background noise. Curves with the same circle marker and triangle marker belong to joint and separate training systems, respectively, and curves with the same colour belong to the two systems with the same bandwidth.

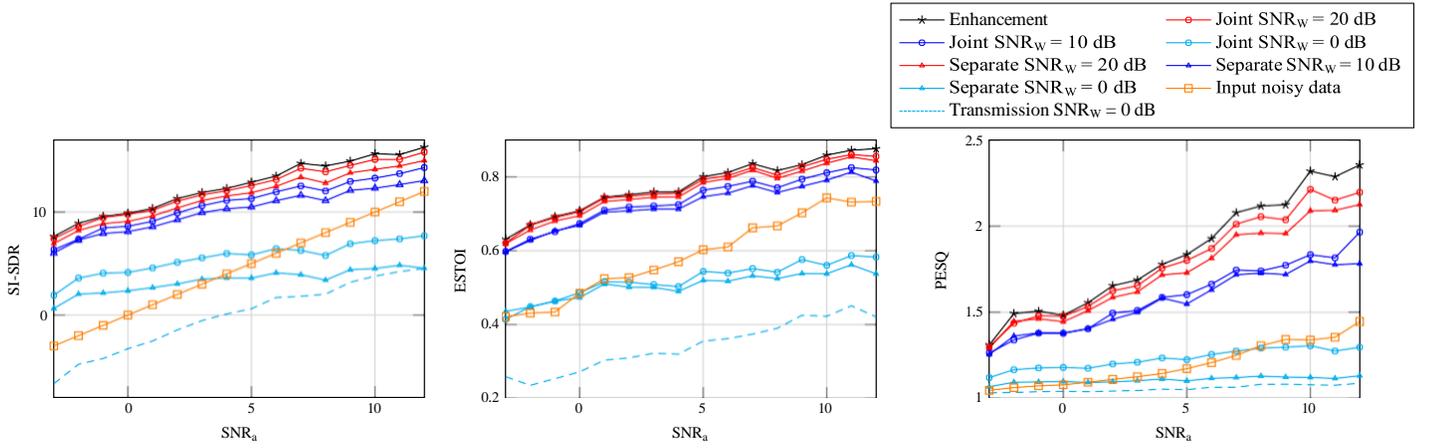

**Fig. 3**: Performance of the proposed joint enhancement and transmission methods for different wireless channel conditions ($SNR_W$) versus SNR of the speech signal against background noise.

method is Conv-TasNet, followed by TransNet (Fig. 1(c.2)). We employ the SI-SDR metric

$$L = \text{SI-SDR}(\mathbf{x}, \hat{\mathbf{x}}) \qquad (1)$$

as the cost function to train the system in an end-to-end manner. This training approach aims to optimize the joint enhancement and transmission performance, enabling the system to effectively address background noise and wireless transmission challenges in real-world scenarios.

## 3. SIMULATION RESULTS

In this section, we present a simulation study to demonstrate the effectiveness of our proposed methods. We evaluate the performance of the proposed approaches under different scenarios, considering speech signals in a range of acoustic signal-to-noise ratios $SNR_a$, various bandwidth compression ratios, diverse wireless transmission channel SNRs ($SNR_W$), different system latencies and system orders. To assess the performance, we employ three key metrics for comparison: Perceptual Evaluation of Speech Quality (PESQ) [23] as a speech quality metric, SI-SDR as a reconstruction metric, and Extended Short-Time Objective Intelligibility (ESTOI) [24] as an intelligibility metric. For training and testing, we use the LibriMix dataset [25], comprising

13900 wav files with a sampling frequency of 16 kHz. The dataset combines the Librispeech dataset [26] for clean speech and the WHAM! noise dataset [27] to create different SNR combinations. In all simulations, $SNR_a$ values are generated by the LibriMix dataset, ranging from −5 dB to 12 dB.

During the training phase, we divide the data into an actual training set (80%) and a validation set (20%). We utilize the Adam optimizer [28] for training all systems. For the separate method, where Conv-TasNet and TransNet are trained separately and independently, we set the learning rates to 0.001 and 0.0001, respectively. For joint training, the learning rate is set to 0.0001. To prevent overfitting, we implement early stopping with a patience of 12 epochs and the batch size is set to 8. We initialize the weights of each network randomly, except for the joint system, where we initialize Conv-TasNet's weights with a pre-trained version. This initialization results in improved performance based on all metrics compared to random initialization. Moreover, we observed no significant difference when initializing TransNet's weights with its pre-trained version.

The focus of simulations is to evaluate the performance of the joint and separate methods. The study refers to [16] as the closest state-of-the-art system to the proposed system. Differences in transmission methods and total latency are highlighted between these two systems, with the proposed system using analog transmission and having a total latency of 3 ms, while the system in [16] simulates digital wireless transmission with packet losses and having total latency of 20 ms. Therefore, we contend that a direct comparison with state-of-the-art systems may not be equitable.

### 3.1. Performance vs. Bandwidth

First, we investigate and compare the performance of the two proposed methods across various transmission bandwidths. We set the wireless channel SNR to $SNR_W = 10$ dB, and the total system latency is 3 ms.

Fig.2 illustrates the results, showcasing three figures that represent the performance metrics: SI-SDR, ESTOI, and PESQ versus acoustic SNRs ($SNR_a$). The black curve in the figures represents the performance of the standalone enhancement system, which acts as an upper bound, signifying the performance under ideal transmission conditions, and the dashed curve represents the performance of the standalone transmission system. The orange curve corresponds to the scores obtained from the noisy input data. Hence, the difference between the orange curve and the other performance curves reflects the level of improvement with respect to the noisy input speech achieved after transmission and enhancement. Also, the difference between the dashed curve and the curves at $k/n = 0.25$ indicates the achieved improvement with respect to the standalone transmitting speech system at $k/n = 0.25$. Across all the considered metrics and bandwidths, the joint method consistently outperforms the separate system. Surprisingly, there are instances where the joint system with lower bandwidth surpasses the separate system with higher bandwidth. Additionally, we observe that the performance gap between the joint and separate systems increases for lower bandwidths and higher $SNR_a$. These results underscore the effectiveness of the joint approach, especially under challenging transmission conditions.

### 3.2. Performance vs. Wireless Channel SNR

In this subsection, our focus is on comparing the performance of joint and separate methods under various wireless transmissions $SNR_W$. We maintain a fixed transmission bandwidth compression ratio of $k/n = 1$ and a latency of 3 ms throughout the experiments. Fig. 3 depicts the simulation results. As before, the black curve signifies the performance of the standalone enhancement system, which serves as an upper bound, representing ideal channel conditions, and the dashed curve represents the performance of the standalone transmission system at specific $SNR_W$s.

For higher wireless transmission $SNR_W$, the performance of both the joint and separate methods approach the black curve, indicating greater performance under better channel conditions. Across all considered wireless channels $SNR_W$, the joint approach consistently outperforms the separate method. Notably, the performance gap between the two methods widens for lower wireless transmission SNRs ($SNR_W$) and higher acoustic SNRs ($SNR_a$). For $SNR_W = 0$ dB, the initial observation reveals that the overall performance falls below that of the noisy input data. However, to ascertain the extent of improvement, a comparison with the performance of the standalone transmission system (dashed curve) is imperative. This comparison unveils improvements across all SNRa values.

Taking into account the results from this and previous simulations, it becomes evident that the joint approach is better equipped to handle more challenging wireless transmission scenarios, particularly those involving lower transmission bandwidths and lower transmission SNRs. This finding highlights the robustness and effectiveness of the joint method in adverse conditions.

Table 1: Performance of proposed joint and separate method for various system latencies and orders.

| Methods | SI-SDR (dB) | ESTOI | PESQ |
|---|---|---|---|
| Joint latency 3 ms | 10.71 | 0.72 | 1.54 |
| Joint latency 5 ms | 10.86 | 0.73 | 1.59 |
| Joint latency 9 ms | 11.22 | 0.74 | 1.62 |
| Separate latency 3 ms | 9.32 | 0.70 | 1.46 |
| Separate latency 5 ms | 9.60 | 0.71 | 1.51 |
| Separate latency 9 ms | 9.74 | 0.72 | 1.55 |
| Joint(TransNet + Conv-TasNet) latency 9 ms | 11.32 | 0.74 | 1.62 |
| Separate (TransNet + Conv-TasNet) latency 9 ms | 9.43 | 0.70 | 1.49 |

### 3.3. Performance vs. Latency

In this subsection, our objective is to compare the performance of the proposed joint and separate approaches for different total system latencies. During this simulation, we use $k/n = 1$ and $SNR_W = 10$ dB and Tab. 1 displays the average performance over all acoustic SNRs ($SNR_a$). Both separate and joint methods demonstrate better performance for higher latencies. Furthermore, we observe that increasing the latency leads to only a slight improvement in performance for both the joint and separate methods. Consistent with the previous findings, the joint method outperforms the separate approach across all system latencies.

### 3.4. Performance vs. different orders

In this subsection, we aim to compare the performance of the proposed joint and separate approaches under varying orders of Conv-TasNet and TransNet. The simulation is conducted with the transmission ratio set at $k/n = 1$ and $SNR_W = 10$ dB. The performance evaluation of the (TransNet + Conv-TasNet) configuration is detailed in the last two rows of Tab. 1 for a latency of 9 ms for joint and separate methods. When considering the joint method with the reversed order, a marginal improvement is observed based on the SI-SDR metric, while the other performance metrics exhibit comparable results. In contrast, the separate method with the reversed order demonstrates inferior performance compared to the normal order. This discrepancy might be attributed to Conv-TasNet not being explicitly trained to address errors introduced by TransNet in the reversed order configuration.

### 4. CONCLUSION

In this study, we proposed novel solutions for low-latency joint speech analogue transmission and enhancement, leveraging deep neural networks. By combining two existing system architectures, Conv-TasNet and TransNet, we effectively addressed background noise, speech coding, and transmission channel effects simultaneously. Comprehensive simulation experiments demonstrate the superiority of the proposed joint approach over a simpler system consisting of separate, independently trained Conv-TasNet and TransNet sub-systems for various transmission bandwidths, wireless SNRs, and latency values. The performance improvements achieved through concatenation and joint training of Conv-TasNet and TransNet modules highlight the potential of our approach. In conclusion, our findings underscore the significance of deep neural networks in tackling the challenges of low-latency joint speech transmission and enhancement. We envision that our work could benefit a wide range of applications that require real-time speech communication and enhancement, like wireless hearing assistive devices.